\documentclass[a4paper,11pt]{article}
\usepackage{aaskaiid}

\newcommand{\hi}{\textsc{Hi}}

\usepackage{xspace}
\usepackage{siunitx}
\usepackage{amsmath}
\usepackage{mathtools}
\usepackage[mathcal]{eucal}
\usepackage[capitalize,noabbrev,nameinlink]{cleveref}
\usepackage{ulem}
\usepackage{orcidlink}

\newcommand{\lya}{Lyman-$\alpha$\xspace}
\newcommand{\sHI}{\ensuremath{{\scriptscriptstyle \hi}}}
\newcommand{\AHI}{{\ensuremath{\mathcal{A}_\sHI}}}
\newcommand{\tcm}{21-cm\xspace}
\newcommand{\OmegaHI}{\ensuremath{\Omega_\sHI}}
\newcommand{\bHI}{\ensuremath{b_\sHI}}

\title{Observational Frontiers in the post-EoR 21-cm Intensity Mapping: Lessons from the SKA Pathfinders}
\ShortTitle{Observational Frontiers in 21-cm IM: SKA Pathfinders}

\ShortName{Elahi et al.} 
\author[1]{Khandakar Md Asif Elahi~\orcidlink{0000-0003-1206-8689}}
\emailAdd{asifelahi999@gmail.com}

\affiliation[1]{Centre for Strings, Gravitation and Cosmology, Department of Physics, Indian Institute of Technology Madras, Chennai 600036, India}

\author[2]{Somnath Bharadwaj}
\affiliation[2]{Department of Physics, Indian Institute of Technology Kharagpur, Kharagpur 721 302, India}

\author[3,4]{Philip Bull~\orcidlink{0000-0001-5668-3101}}
\affiliation[3]{Jodrell Bank Centre for Astrophysics, University of Manchester, Manchester, M13 9PL, United Kingdom}
\affiliation[4]{Department of Physics and Astronomy, University of Western Cape, Cape Town 7535, South Africa}

\author[5,6,7,4]{Stefano Camera~\orcidlink{0000-0003-3399-3574}}
\affiliation[5]{Dipartimento di Fisica, Universit\`a degli Studi di Torino, Via P.\ Giuria 1, 10125 Torino, Italy}
\affiliation[6]{INFN -- Istituto Nazionale di Fisica Nucleare, Sezione di Torino, Via P.\ Giuria 1, 10125 Torino, Italy}
\affiliation[7]{INAF -- Istituto Nazionale di Astrofisica, Osservatorio Astrofisico di Torino, Strada Osservatorio 20, 10025 Pino Torinese, Italy}

\author[8,9]{Arnab Chakraborty}
\affiliation[8]{Trottier Space Institute, McGill University, Montreal, QC, Canada}
\affiliation[9]{Department of Physics, McGill University, Montreal, QC, Canada}

\author[10, 11]{Tzu-Ching Chang}
\affiliation[10]{Department of Physics, California Institute of Technology, 1200 E. California Boulevard, Pasadena, CA 91125, USA}
\affiliation[11]{Jet Propulsion Laboratory, California Institute of Technology, 4800 Oak Grove Drive, Pasadena, CA 91109, USA}

\author[12,13,14,15]{Xuelei Chen}
\affiliation[12]{National Astronomical Observatories, Chinese Academy of Sciences, Beijing 100101, China}
\affiliation[13]{School of Astronomy and Space Science, University of Chinese Academy of Sciences, Beijing 100049, China}
\affiliation[14]{State Key Laboratory of Radio Astronomy and Technology, Beijing 100101, China}
\affiliation[15]{Key Laboratory of Cosmology and Astrophysics (Liaoning) \& College of Sciences, Northeastern University, Shenyang 110819, China}

\author[1]{Samir Choudhuri~\orcidlink{0000-0002-2338-935X}}
\author[16]{Devin Crichton~\orcidlink{0000-0003-1204-3035}}
\affiliation[16]{Institute for Particle Physics and Astrophysics, Department of Physics, ETH Zurich, Wolfgang-Pauli-Strasse 27, 8093 Zurich, Switzerland}

\author[17]{Abhirup Datta~\orcidlink{0000-0002-5333-1095}}
\affiliation[17]{Department of Astronomy, Astrophysics and Space Engineering, Indian Institute of Technology Indore, Indore 453552, India}

\author[18]{Simon Foreman}
\affiliation[18]{Department of Physics, Arizona State University, Tempe, AZ, USA}

\author[4,12]{Wenkai Hu~\orcidlink{0000-0002-3108-5591}}

\author[15]{Yichao Li} 

\author[19, 20]{Kiyo Masui}
\affiliation[19]{MIT Kavli Institute for Astrophysics and Space Research, Massachusetts Institute of Technology, Cambridge, MA, USA}
\affiliation[20]{Department of Physics, Massachusetts Institute of Technology, Cambridge, MA, USA}

\author[3]{Ainulnabilah Nasirudin}
\author[17]{Samit Kumar Pal~\orcidlink{0000-0002-2271-4165}}
\author[17]{Rashmi Sagar}

\author[8,9,21]{Seth R. Siegel}
\affiliation[21]{Perimeter Institute for Theoretical Physics, Waterloo, ON, Canada}

\author[22]{Lister Staveley-Smith}
\affiliation[22]{International Centre for Radio Astronomy Research (ICRAR), University of Western Australia, Crawley, WA 6009, Australia}

\author[23]{Eric Switzer}
\affiliation[23]{NASA Goddard Space Flight Center, 8800 Greenbelt Road, Greenbelt, MD 20771, USA}

\author[12,13,14,15]{Yougang Wang}
\author[3]{Laura Wolz}
\author[12,13]{Wenxiu Yang}
\author[12,14]{Shifan Zuo}

\abstract{
The 21-cm line from neutral hydrogen has long been recognised as a promising tracer of the large-scale structure of the Universe. The line is weak however, making individual galaxy detections quite inefficient, especially at higher redshifts. The technique of 21-cm intensity mapping has been pioneered over the last two decades to address this limitation. Instead of detecting individual galaxies, the brightness temperature field from the combined 21-cm emission of many unresolved galaxies is mapped as a function of angle and frequency, resulting in 3D tracer maps of the large-scale structure. In this chapter, we review the major pioneering efforts to develop this observable into a competitive cosmological tool, paying particular attention to the status of pathfinder observations that have paved the way for a large and highly sensitive 21-cm intensity mapping survey with the SKA-Mid telescope. 

\vspace{1cm}
Additional keywords: Intensity Mapping, Precursors/pathfinder results
}

\begin{document}
\maketitle

\section{Introduction}

The history of the SKA Observatory (SKAO) is intimately connected with the desire to observe neutral hydrogen (\hi) out to high redshifts \citep{2012arXiv1212.3497E}. The 21-cm line from \hi\ is a faint but omnipresent tracer of galaxies in the radio part of the spectrum. The vast majority of galaxies of any type contain large reservoirs of neutral gas that are sufficiently dense as to be self-shielded from ionising UV photons. Given that the 21-cm line is in an isolated part of the spectrum (making interloper detections unlikely) and in principle detectable across a very wide redshift range, it should be an excellent tracer of the cosmic large-scale structure. However, the faintness of the line calls for very deep, high spectral resolution observations to detect it, hence early calculations posited that around a square kilometre of collecting area would be required to observe the majority of galaxies through this line out to redshifts far beyond the $z \approx 0.1-0.2$, which were being probed by Arecibo at the time \citep{1991ASPC...19..428W}.

Detecting individual galaxies is an inherently inefficient process. Candidate line detections must be clearly above the noise to avoid the risk of large numbers of spurious detections entering the resulting galaxy catalogue. Thresholding at several times the noise level---typically $5$ to $10\,\sigma$---will make spurious detections rare, but also throws away a great many detections of faint galaxies that are above, but not sufficiently above, the noise level. In other words, the majority of the detected radio emission is thrown away, leaving only the brightest `peaks'. For the 21-cm line, it is also common to apply a filter or cut on line width to help reject spurious bright lines from radio frequency interference for example. Galaxies that are side-on to the observer then present a characteristic ``double-horned'' line profile that is indisputable as a detection. On the other hand, a large fraction of face-on galaxies are excluded by this filter.

The intensity mapping (IM) technique has been developed to side-step these issues. Instead of catalogues of galaxies, this observable takes the brightness temperature field as a whole, constructing a 3D map as a function of direction and frequency that contains the 21-cm emission from {\it all} of the galaxies within the survey volume. The 21-cm field can be separated from noise and other effects on a statistical basis; for sufficiently low-noise (and efficiently calibrated and cleaned) data, the actual large-scale structure field can become visible as an actual 3D map, but this is not necessary. Instead, the cosmological signal can be extracted through two- and three-point correlation statistics, such as the power spectrum and bispectrum in Fourier space, if the noise level is low enough on the scales (Fourier wavenumbers) of interest. Current attempts to detect the 21-cm brightness temperature fluctuations largely focus on statistical detection of this kind, as most of the available cosmological information can be extracted from the power spectrum long before the signal emerges in map space.

As outlined in other chapters, there are several challenging aspects to extracting the 21-cm signal, which remains faint (i.e.\ an rms at the mK level) even through the IM technique. Galactic and extragalactic radio emission, particularly synchrotron emission, is several orders of magnitude brighter, and must be removed efficiently to reveal the cosmological 21-cm signal. The spectral smoothness of this `foreground' emission provides an important distinguishing feature from the 21-cm signal, allowing a clean separation in principle, but this is further complicated by calibration and instrumental systematics, which modulate the signals. Because of the high dynamic range between the foreground and 21-cm signals, even a slight miscalibration or errors in the foreground removal procedure can result in unacceptable levels of residual contamination. For these reasons, so much of the last two decades of 21-cm cosmology have been focused on improving calibration and foreground removal techniques, and handling the additional problems that these introduce, such as signal loss---unavoidable over-fitting of foregrounds that also suppresses the 21-cm signal.

Great progress has been made on these problems however, and several statistical detections of the cosmological 21-cm signal have been made. Most of these rely on cross-correlations with optical galaxy surveys \citep{2010Natur.466..463C,2013ApJ...763L..20M,2013MNRAS.434L..46S,2017MNRAS.464.4938W,2018MNRAS.476.3382A,2022MNRAS.510.3495W,2023ApJ...947...16A,2024ApJ...963...23A,2023MNRAS.518.6262C}, which have independent systematic effects; the residual foreground and systematic contaminants in the 21-cm data essentially correlate out, leaving only the true cosmological cross-correlation signal up to some undetermined amplitude factor (the correlation coefficient between the 21-cm field and the galaxy field). 
Attempts at direct (auto-correlation) detections have also been made.

In what follows, we review the status of 21-cm IM observations on a number of SKAO pathfinder telescopes\footnote{A list of the SKAO pathfinders and precursors can be found here: \url{https://www.skao.int/en/explore/precursors-pathfinders}.} arranged in alphabetical order by name. In addition to providing several scientifically valuable detections, these experiments have substantially contributed to the development of the calibration and data analysis methods necessary to make large-scale 21-cm IM surveys on SKA-Mid a reality. The rest of the chapter is arranged as follows: \cref{sec:chime} describes CHIME\footnote{\url{https://chime-experiment.ca/}}'s measurements of the 21-cm IM signal using cross-correlation techniques. \cref{sec:fast} presents the ongoing and upcoming surveys and forecasts of FAST\textbf{\footnote{\url{https://fast.bao.ac.cn}}}. \cref{sec:gbt} describes several cross-power spectrum detection of the \hi{} signal with GBT and the main lessons for future surveys. \cref{sec:gmrt} assimilates the effort of GMRT\footnote{\url{http://www.gmrt.ncra.tifr.res.in/}} towards measuring the signal in auto-correlation and the possibilities of using similar techniques for SKA-Mid. The early deployment and systematic mitigation strategies of HIRAX, as it is being designed and commissioned, are presented in \cref{sec:hirax}. \cref{sec:Murriyang} describes the measurements with Murriyang (Parkes\footnote{\url{https://www.atnf.csiro.au/facilities/murriyang-our-parkes-radio-telescope}}) and its future goals of testing a new cryogenic phased array feed for IM surveys. 
\cref{sec:tianlai} presents the cylinder and dish pathfinder arrays of the Tianlai project\footnote{ \url{https://tianlai.bao.ac.cn}.}, with a description of their present status and future goals. \cref{sec:Conclusions} concludes the chapter. Additionally, we refer the reader to \cite{Cunnington01.2026.SKA}, a separate chapter on pathfinder observations with MeerKAT, due to its role as a direct precursor to SKA-Mid.

\section{CHIME} 
\label{sec:chime}

The Canadian Hydrogen Intensity Mapping Experiment (CHIME) is a transit radio telescope located at the Dominion Radio Astrophysical Observatory near Penticton, British Columbia, Canada. The instrument is designed to map \hi{} over the redshift range $0.8-2.5$ to constrain the expansion history of the universe through 21-cm IM. CHIME consists of four parallel cylindrical reﬂectors, oriented north–south, each 100~m $\times$ 20~m and instrumented with 256-element dual-polarization cloverleaf antennas \citep{Deng2017arXiv170808521D} observing at $400-800$~MHz. CHIME observes a two-degree-wide stripe covering the entire meridian at any given moment, observing three-quarters of the sky every day owing to Earth’s rotation. Signals from each feed are processed by an FX correlator and stored for ofﬂine cosmological analysis with a frequency resolution of 390~kHz. These signals are also fed to separate back ends devoted to studying fast radio bursts \citep{FRB2018}, pulsars \citep{CHIMEPulsar:2021}, Galactic magnetism \citep{chime-tadpole}, and \tcm absorption systems \citep{chime-absorber}. \citet{chimeoverview} describes the telescope design and subsystems in more detail.

\begin{figure*}
   \centering \includegraphics[width=0.98\linewidth,keepaspectratio]{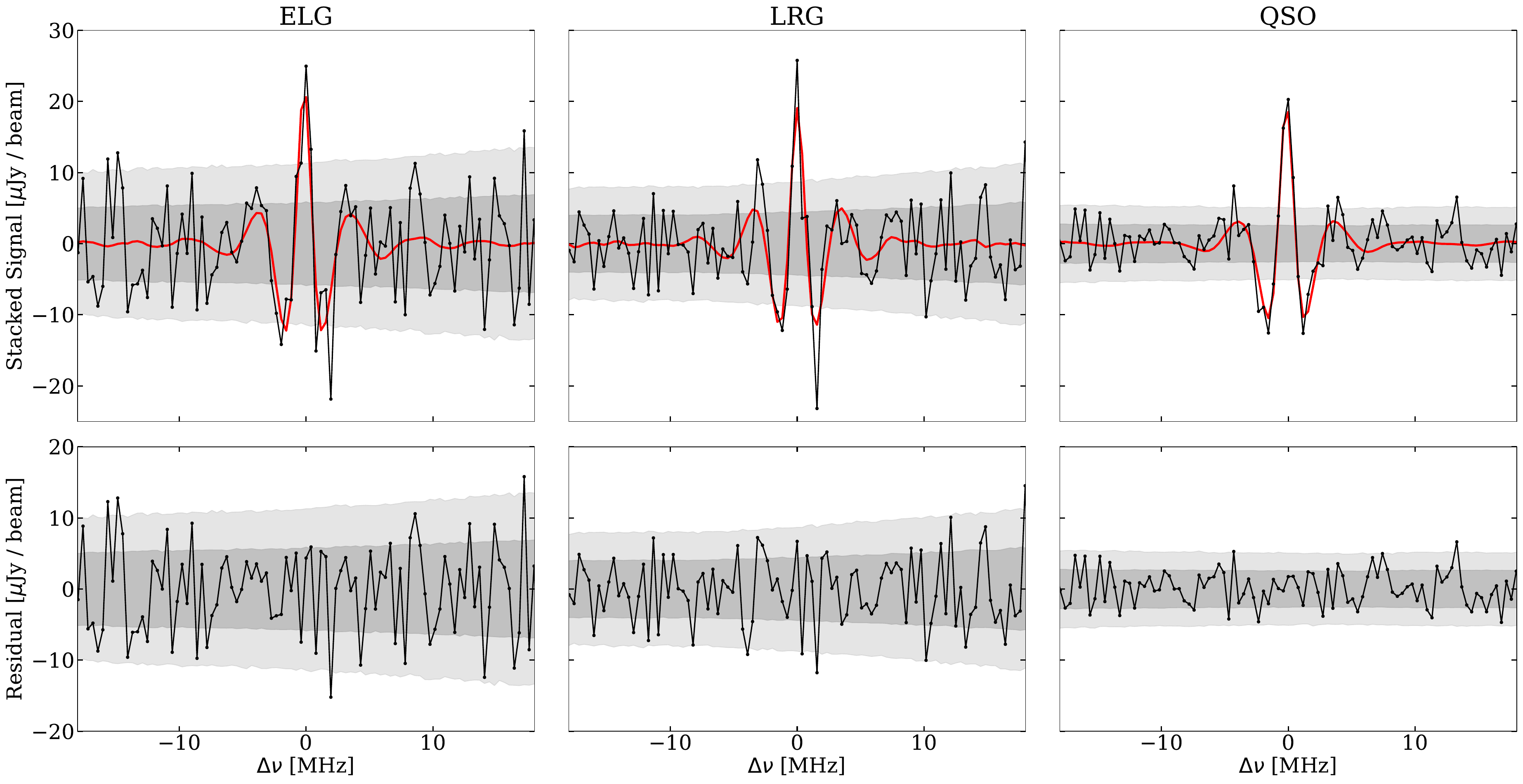}
    \caption{\emph{(Top)} The CHIME stacking signal as a function of frequency offset for the ELG, LRG, and QSO catalogs.  The data are shown in black and the best-fit model is shown in red.  \emph{(Bottom)} The residuals obtained by subtracting the best-fit model from the data.  For both the top and bottom rows, the dark gray and gray bands indicate the central \SI{68}{\percent} and \SI{95}{\percent} of values observed when applying the same stacking procedure to \SI{10000} mock catalogs.}
    \label{fig:stack1d}
\end{figure*}

In \citet{chimestacking}, we report CHIME's first detection of the 21-cm emission from large-scale structure (LSS) between redshift 0.78 and 1.43 using 102 nights of CHIME data acquired in 2019. 
We construct a confusion noise dominated map of the northern sky, which is foreground filtered  using a linear high-pass filter. We perform spectral stacking of the filtered map on the angular and spectral locations of luminous red galaxies (LRG), emission line galaxies (ELG), and quasars (QSO) from the eBOSS clustering catalogs \citep{dawson2016}, yields a detection significance of $7.1\sigma$ (LRG), $5.7\sigma$ (ELG), and $11.1\sigma$ (QSO). These are the first 21-cm IM measurements made with an interferometer in cross-correlation at this redshift range. 
\Cref{fig:stack1d} shows the central pixel of the stack as a function of frequency offset, i.e., $d(\Delta \nu, 0, 0)$, for the three different tracers in black.  The dark gray and light gray contours indicate the central \SI{68}{\percent} and \SI{95}{\percent} of values observed when stacking the maps on \SI{10000}{} random mock catalogs as outlined.  The red line indicates our best-fit model for the signal.

We interpret these measurements using a simulation-based framework, within a model that considers \hi{} and galaxies to be linearly biased tracers of the underlying matter distribution, including the leading effects of redshift-space distortions and a correlated shot noise contribution related to the mean \hi{} mass of the objects in each catalog. 
We constrain the effective clustering amplitude of \hi{}, defined as $\AHI \equiv 10^3 \, \OmegaHI (\bHI + \langle f\mu^2 \rangle)$, where $\OmegaHI$ is the cosmic abundance of \hi{}, $\bHI$ is the linear bias of \hi{}, and $\langle\,f\mu^{2}\rangle=0.552$ encodes the effect of redshift-space distortions at linear order.  We constrain this amplitude separately for each eBOSS catalog, marginalizing over parameters controlling the scale dependence of non-linear clustering, obtaining $\AHI=1.51_{-0.97}^{+3.60}$ (LRGs) $(z=0.84)$, $\AHI=6.76_{-3.79}^{+9.04}$ (ELGs) $(z=0.96)$, and $\AHI=1.68_{-0.67}^{+1.10}$ (QSOs) $(z=1.20)$. These constraints are limited by modeling
uncertainties at nonlinear scales. We also split the QSO catalog into three redshift bins and have a decisive detection in each, with the upper bin at $z=1.30$ producing the highest redshift 21-cm IM measurement thus far.

In \citet{chime-lymanalpha}, we extended our analysis to higher redshifts and report the detection of \tcm signal at an average redshift $\bar{z} =2.3$ in the cross-correlation of CHIME with measurements of the \lya forest from eBOSS.  This analysis used 88 days of CHIME data in the 400-500 MHz frequency band, corresponding to redshifts $1.8 < z < 2.5$. The \lya forest is characterized by absorption of quasar light by \hi{} clouds along the line of sight, whereas \tcm signal is in emission. Hence we expect the cross-correlation of \lya forest with CHIME maps to be negative at small separations, which provides a distinctive signal that is difficult to mimic by systematic effects. The analysis extracted line-of-sight spectra from foreground-filtered CHIME maps at the locations of background quasars from the eBOSS DR16 catalog \citep{de_Mas_2020} and combined them with the \lya forest flux transmission spectra to estimate the cross-correlation function.

\begin{figure*}[htb]
    \centering \includegraphics[width=0.98\linewidth,keepaspectratio]{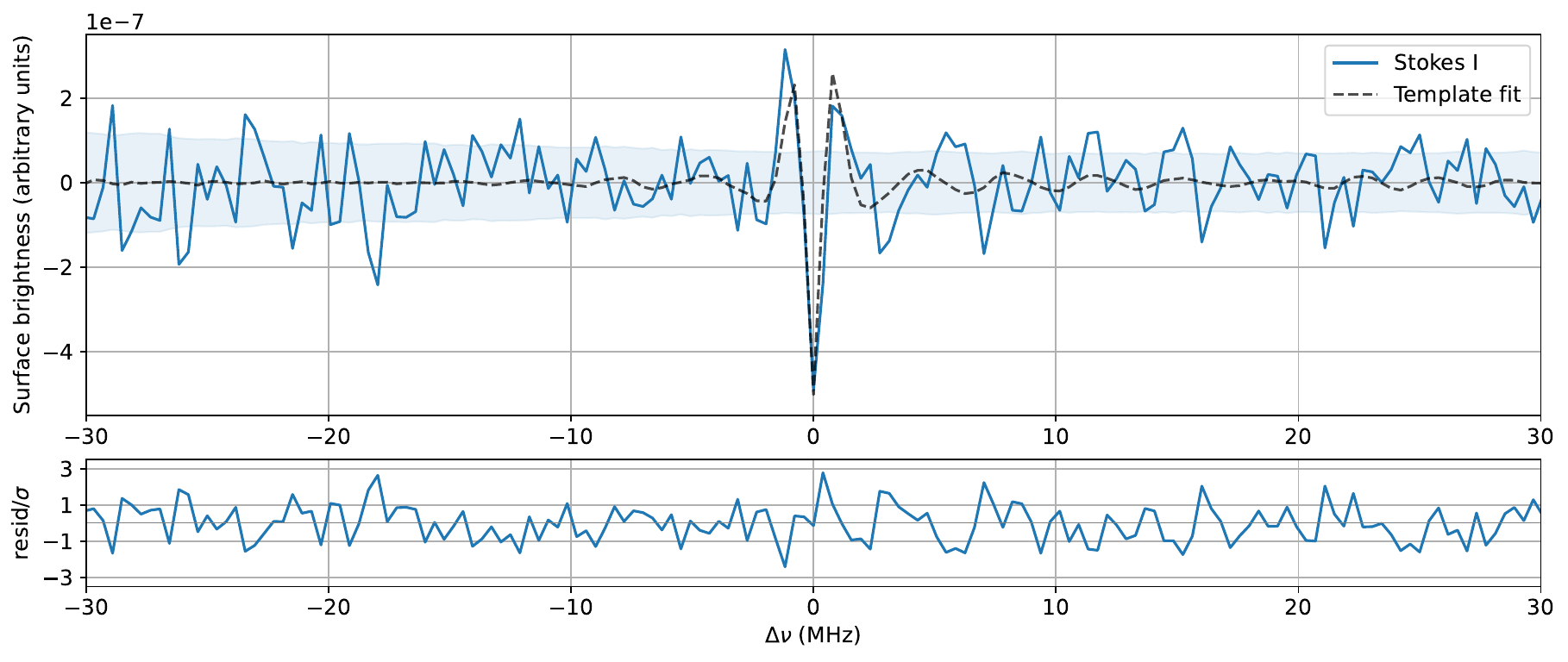}
    \caption{\emph{(Top)} Cross-correlation function of CHIME and eBOSS \lya
    forest data, for the combined X and Y polarisations. An estimate of the
    standard deviation of the background is plotted as a shaded region. The black dashed
    line is a template derived from simulations with an amplitude fit to the
    data. \emph{(Bottom)} Residuals normalised by the estimated background
    level.}
    \label{fig:xcorr_fit}
\end{figure*}

The cross-correlation functions of CHIME and the eBOSS \lya forest are presented
in \cref{fig:xcorr_fit}. We fit a simulation-derived template function to this measurement results, yields a $9\sigma$ detection significance. These results are  the highest-redshift measurement of \tcm emission to date, and set the stage for future \tcm IM analyses at $z>1.8$.

In both analyses, the foreground filtering removes sensitivity to linear cosmological scales related to Baryon Acoustic Oscillations (BAOs), but enables high signal-to-noise-ratio (SNR) measurements at non-linear scales. These detection in cross-correlations establishes CHIME's capability for precision 21-cm cosmology and opens multiple avenues for future investigation. These results only use less than 1 year of observations, and the CHIME archive now contains nearly 6 years of data. The CHIME collaboration is working on analyzing this data, and will present the results in future work. The collaboration is also actively working on measuring the \tcm signal as an auto-power spectrum at non-linear scales. In parallel, work is ongoing on improvements of our beam model and exploring new foreground filtering methods to probe the linear BAO scale, thereby providing powerful constraints on dark energy and the universe’s expansion history.

\section{FAST}
\label{sec:fast}

The Five-hundred-meter Aperture Spherical radio Telescope \citep[FAST,][]{2011IJMPD..20..989N,2016RaSc...51.1060L} is the largest single-dish radio telescope in the world today. 
It is located in Dawodang karst depression, a natural basin in Guizhou province, China
$({\rm E}\,106\overset{\circ}{.}86,\,{\rm N}\,25\overset{\circ}{.}65)$. 
With a maximum zenith angle of $40^{\circ}$, FAST can observe sources at declinations between approximately $-15^\circ$ and $65^\circ$, covering a sky area about $25\,000\, \deg^2$. 
FAST is currently equipped with the L-band ($1.05 - 1.45$ GHz) 19-beam receiver \citep{2020RAA....20...64J}, corresponding to the \hi{} redshift range of $0 < z < 0.35$. The illuminated aperture during observations is 300 meters in diameter, resulting in a geometric collecting area of about $70700 \,\rm m^2$. 
Additionally, with $\sim 2.95\,{\rm arcmin}$ angular resolution at 1420~MHz for each beam, the large area surveys with FAST can potentially probe low-mass \hi{} galaxies (even the dark galaxies) and the \hi{} distribution in a large number of galaxy clusters, groups, filaments, and voids.

\subsection{Current surveys and status}

FAST has been conducting several large area survey programs, including both dedicated experiments for \hi{} IM and general-purpose astronomical observations. For example, the FAst neuTral HydrOgen intensity Mapping ExpeRiment \citep[FATHOMER,][]{2023ApJ...954..139L} is a pilot survey for \hi{} IM with FAST. 
The primary scientific goals of the project are to study the large-scale structure of the Universe, galaxy formation and evolution through wide-area mapping of the \hi{} signal.
Since the start of observations in 2020 \citep{2021MNRAS.508.2897H}, FATHOMER has covered approximately 150 square degrees of the sky near the zenith of FAST using the drift-scan mode. 
A complete data processing pipeline ({\tt fpipe}\footnote{\url{https://github.com/TianlaiProject/fpipe}}) for drift scan surveys with FAST has been made publicly available, performing the main processing steps from raw time-ordered data to science-ready map cubes. With this pipeline, we have produced well-calibrated maps for a region of about $60\, \deg^2$. Preliminary results for continuum point source detections in this region have also been released in \cite{2023ApJ...954..139L}.
These data are currently being used for the analysis of systematic effects (e.g., beam construction by \citealt{2025AJ....169..265Z}), a blind \hi{} galaxy survey, and preliminary estimation of the \hi{} power spectrum.

Data from other large surveys, like the Commensal Radio Astronomy FAST Survey \citep[CRAFTS,][]{2018IMMag..19..112L} and the FAST All Sky \hi{} survey \citep[FASHI,][]{2024SCPMA..6719511Z}, will also be explored for \hi{} IM studies in the future. In particular, CRAFTS is a drift scan survey project designed to simultaneously observe transients such as pulsars and fast radio bursts (FRBs), as well as conduct spectral surveys including \hi{}. For this purpose, it employs a novel high-cadence calibration mode and uses both the pulsar backend and the spectral backend to record data. This project plans to scan the whole sky accessible to FAST twice in the next few decades. Since its start in 2020, it has covered an area of $\sim$~9000~$\mathrm{deg^2}$, distributed widely across the sky. Like FATHOMER, we have also developed the data processing pipeline for CRAFTS based on {\tt fpipe} and performed validation tests for continuum sources and \hi{} galaxies detection \citep{2025ApJS..279...32Y}. The large sky coverage of CRAFTS provides substantial opportunities for \hi{} IM studies. 

\subsection{Forecast and future plan}

Forecasts in \cite{2020MNRAS.493.5854H} indicate that a FAST drift-scan survey with the L-band 19-beam receiver, covering the FAST-accessible sky ($\sim 20~000\,{\rm deg}^2$) in about 220 days, would allow a good detection of the \hi{} power spectrum. It is expected to achieve an 
SNR~$> 5$ at redshift $0.05 < z < 0.35$ on the BAO scale ($\sim 0.1 h/{\rm Mpc}$), through both \hi{} IM and \hi{} galaxy survey. Generally, the \hi{} IM technique will perform better than the HI galaxy survey in terms of \hi{} power spectrum precision, especially at high redshift. 

An ultra-wide bandwidth receiver at $500-3300$ MHz is under testing and planned for future deployment \citep{2023RAA....23g5016Z}. Moreover, the FAST Core Array has been proposed as an extension, which will integrate FAST into a synthesis aperture array comprising 24 secondary 40-m antennas located within 5 km of the FAST site \citep{2024AstTI...1...84J}. With the upgraded instrumentation, it will be possible to probe the Universe at higher redshifts and multiple scales with FAST. 

Looking ahead, as deeper and wider observations accumulate, along with continuous improvements in data processing techniques and a better understanding of system performance, FAST is expected to play an important role in \hi{} IM in the near future. It will also provide valuable precursor measurements that can complement upcoming surveys with the SKA.

\section{Green Bank Telescope}
\label{sec:gbt}

The Green Bank Telescope (GBT) has been the pioneering radio facility enabling the first detection of the \hi\ IM signal as proposed in \cite{2008PhRvL.100i1303C}. The Green Bank Telescope is a fully-steerable 100~m diameter dish radio telescope based in a radio-quiet area in Green Bank, West Virginia, US. The system temperature of the single pixel feed is 25K for frequencies below 1000 MHz. 
\begin{figure}
    \centering
    \includegraphics[width=0.68\linewidth]{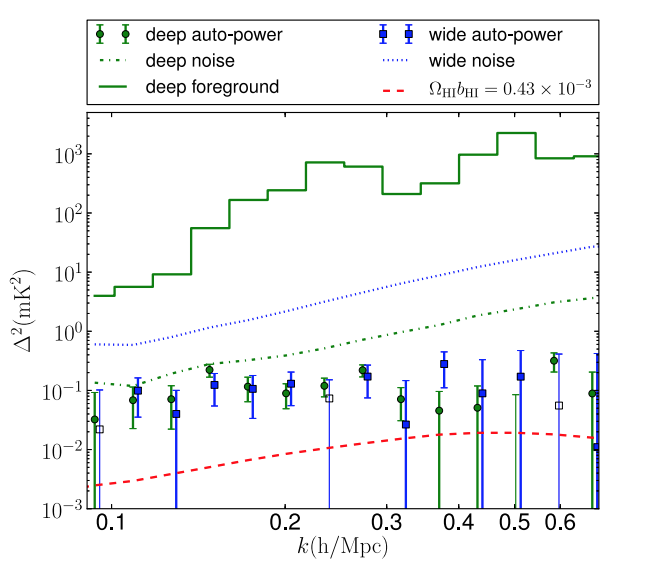}
    \caption{The solid curve is the power spectrum of the input deep field with no cleaning applied. Throughout, the deep field results are green and the wide field results are blue. The dotted and dash–dotted lines show thermal noise in the maps. The power spectra avoid noise bias by crossing two maps made with separate data sets. The points below show the power spectrum of the deep and wide fields after the foreground cleaning. Individual modes in the map are dominated by thermal noise rather than residual foregrounds or signal. The negative values are shown with thin lines and hollow markers. The red dashed line shows the expected 21-cm auto-power spectrum obtained from the amplitude of the WiggleZ galaxy -- \hi{}  cross-power with $r_{\rm cross}=1$ ($r_{\rm cross}$ is the WiggleZ galaxy--\hi{} cross-correlation coefficient), and based on simulations processed by the same pipeline. Figure taken from \cite{2013MNRAS.434L..46S}.    }
    \label{fig:gbtPS}
\end{figure}

The first detection of the large-scale structure within the \hi{} IM signal  at $z \approx 1$ was reported in \cite{2010Natur.466..463C} using 15 hr of integration time on a 2 sq deg in cross-correlation with the DEEP2 survey. Consequently, based on the initial detection, a larger survey with the 680-920 MHz prime-focus receiver was designed observing multiple fields with complete overlap with spectroscopic galaxy sample of the same redshift range $0.6<z<1$.
In \cite{2013ApJ...763L..20M} and \cite{2013MNRAS.434L..46S} the first cross-correlation detection of the \hi{} signal with WiggleZ galaxies was reported using 105 hr of integration on the `15 hr deep field'  and 84 hrs of integration the `1 hr wide field'. The observing strategy of the survey was a drift scan in azimuthal scans which created a range of crossing angles between scans and an isotropic noise within the centres of the patches. The final analysis used 200 MHz bandwidth of data with 4096 channels with only minimal loss of frequencies due to two telescope resonances at 798 MHz and 817 MHz. After calibration and map-making, the data was processed in square pixelisation of 0.0627 deg angular size and 0.78 MHz frequency width. The telescope beam evolves between 0.25 deg and 0.31 deg at frequencies $(700<\nu <900)$ MHz.

\begin{figure}
    \centering
    \includegraphics[width=0.45\linewidth]{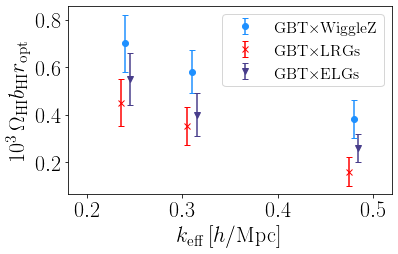}
    \includegraphics[width=0.45\linewidth]{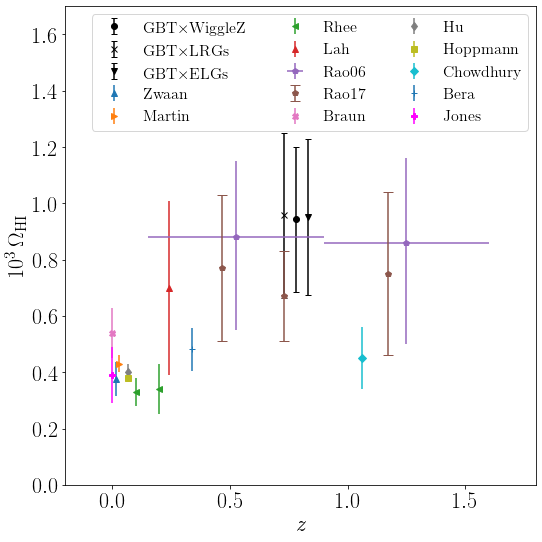}
    \caption{\textit{Left panel:} Best-fit and 1$\sigma$ statistical errors on $10^3\Omega_\hi b_\hi r_{\rm cross}$ at a mean redshift $z = 0.78$ for $N_{\rm IC} = 36$ (number of independent component removed in the foreground separation), together with the effective scale $k_{\rm eff}$ (staggered for illustration purposes). \textit{Right panel:} Estimates for HI from GBT IM compared to other measurements in the literature. All our estimates are at the central redshift z = 0.78 but they have been staggered for illustration purposes. We used the results $k_{\rm eff} = 0.24\rm \ h Mpc^{-1}$ for deriving these estimates. Figures taken from \cite{2022MNRAS.510.3495W}.}
    \label{fig:gbtomHI}
\end{figure}
\cite{2013ApJ...763L..20M} demonstrate a cross-power spectrum detection of the \hi{} signal in both fields, which 
\cite{2013MNRAS.434L..46S} combined with upper bounds from auto-power to obtain a measurement of $\Omega_\hi b_\hi =(0.62^{+0.23}_{-0.15})\times 10^{-3}$.
\autoref{fig:gbtPS} shows the detected power spectrum signal, both the upper limit from the auto-correlation and the WiggleZ cross-correlation, where blue and green lines show results from the `wide' and `deep' field, respectively.

In \cite{2022MNRAS.510.3495W}, an extended version of the 1 hr field of the GBT survey was analysed with an extended set of spectroscopic samples using WiggleZ Dark Energy Survey as well as the SDSS Emission Line Galaxy (ELG) and Luminous Red Galaxy (LRG) sample of the eBOSS survey (DR16). The 1 hr field was extended to 100 sq deg using 100 hr of integration. The analysis was performed using an independent foreground removal method and power spectrum analysis pipeline which had been tested on the original data in \cite{2017MNRAS.464.4938W}. Signal loss due to foreground separation was compensated via a transfer function as in previous analysis \cite{2013MNRAS.434L..46S}, an extended formalism of which is presented in \cite{2015ApJ...815...51S}. The data analysis was accompanied by simulations of the correlation using post-processed semi-analytic simulations based on \cite{2016ApJS..222...22C}, in order to predict the cross-correlation coefficient between \hi{} and individual galaxy samples. The analysis shows that the cross-power amplitude $\Omega_\hi b_\hi r_{\rm cross}$ is moderately dependent on foreground removal parameters and strongly dependent on the fitted $k$ range. This is due to non-linearities of the amplitude parameters $b_\hi$ as well as $r_{\rm cross}$. The constraints on $\Omega_\hi b_\hi$ using predicted $r_{\rm cross}$ are shown in \autoref{fig:gbtomHI}.

\textbf{Key Lessons Learned}
\begin{itemize}
    \item Division of data into independent time blocks for internal cross-correlations is essential for systematic check and understanding noise characteristics.
    \item Most foreground removal methods perform equivalently; the major limitation is initial map quality.
    \item Characterizing signal loss during foreground removal is critical, where non-linear algorithms may introduce subtle effects that require careful thought.
\end{itemize}

\section{GMRT}
\label{sec:gmrt}


The first proposal to map the LSS using the redshifted 21-cm  line was presented in \cite{Bharadwaj2001a}, who demonstrated both the mean and the power spectrum of the \hi{} brightness temperature fluctuations can be detected by the Giant Metrewave Radio Telescope (GMRT; \citealt{Swarup1991}), the largest telescope operating at meter wavelengths at that time. 
GMRT has 30 fully steerable, 45~m diameter parabolic dish antennas, arranged in a `Y-shaped' configuration spanning approximately 25~km. The total collecting area is about 30~000~m$^2$ at metre wavelengths, with an angular resolution of a few arcseconds. The  recently upgraded GMRT (uGMRT; \citealt{Gupta2017}) has nearly seamless frequency coverage over the range 120 -- 1500 MHz with large instantaneous bandwidths (Band 2: 120 -- 250 MHz; Band 3: 300 -- 500 MHz, Band 4: 550 -- 850 MHz, Band 5: 1050 -- 1450 MHz), a few kHzs of frequency resolution, and an improved receiver, which makes it the most sensitive interferometer over a wide radio band prior to the SKA (e.g., Fig. 8 of \citealt{braun2019anticipatedperformancesquarekilometre}).

\subsection{GMRT Observations for 21-cm IM:} 

The first attempt at post-reionization 21-cm IM with GMRT was carried out by \cite{2011MNRAS.411.2426G} at 610~MHz ($z=1.3$), who identified that the sidelobe response of the telescope causes wide-field foregrounds to appear as an oscillatory pattern in the measured $C_\ell(\Delta\nu)$ the multi-frequency angular power spectrum (MAPS).  
\cite{2011MNRAS.418.2584G} (later \citealt{samir14, samir16}) showed  that the amplitude of the oscillations can be suppressed by tapering the primary beam response of the telescope, and they placed an upper limit of $[\Omega_{\rm \hi} b_{\rm \hi}] < 0.11$ using the measured $C_\ell(\Delta\nu)$.

More recently, uGMRT's Band 3 was used to perform a detailed analysis of the ELAIS-N1 field (\citealt{Arnab2019A, Arnab2019B, Arnab2020}). The observation has since facilitated several progressively tighter upper limits on the \hi{} IM signal, starting with a multi-redshift constraints at $1.96 < z  < 3.58$ \citep{Chakraborty2021}.  
On the same data, \cite{2022MNRAS.516.2851P} used the estimator TGE, in which, visibilities are correlated to estimate the $C_\ell(\Delta\nu)$, which are Fourier transformed to estimate the cylindrical power spectrum $P(k_\perp, k_\parallel)$, thereby taking care of the missing frequency channels \textit{before} moving to the Fourier space \citep{Bh18}. 
They demonstrated that wide-field foreground can be efficiently mitigated by tapering the beam response, and denser `$uv$'-coverage further improves this aspect. 
\cite{2023MNRAS.520.2094E} then improved the upper limits by cross-correlating two mutually orthogonal polarizations while estimating $C_\ell(\Delta\nu)$. The cross-correlation mitigated several systematics, and the estimated power spectrum was found to be consistent with noise.

\begin{figure}
    \centering
    \includegraphics[width=0.95\linewidth]{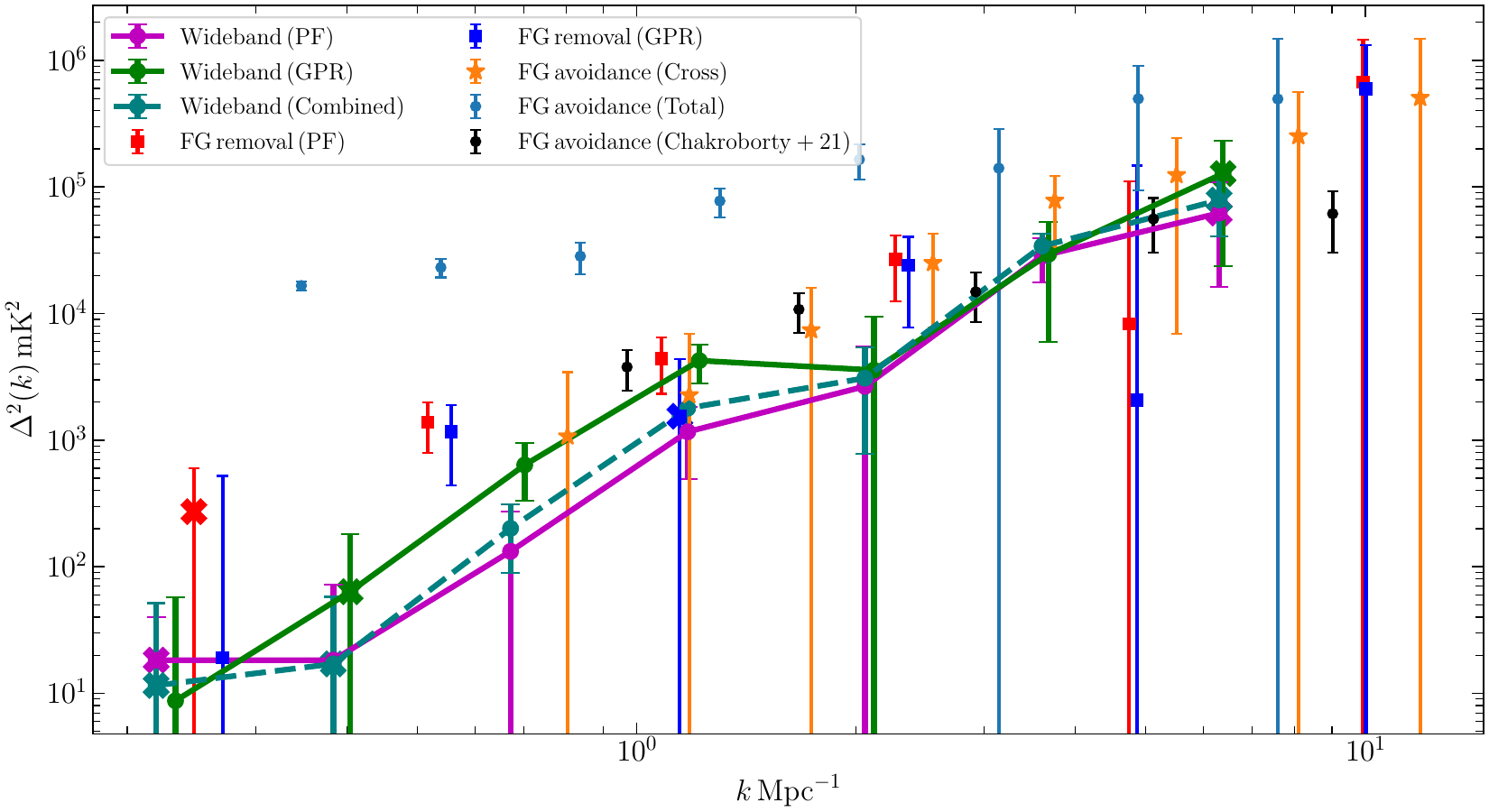}
    \caption{The measured $\Delta^2(k)$ and  $2\sigma$ uncertainties obtained from uGMRT Band 3 observations. The three curves show the results from \cite{2024MNRAS.529.3372E} using different methods as labelled and described in the text. The other data points from \cite{2023MNRAS.525.3439E} (squares), \cite{2023MNRAS.520.2094E} (asterisks), \cite{2022MNRAS.516.2851P} (light blue circles), and \cite{Chakraborty2021} (black circles) are shown to show systematic improvements towards 21-cm IM with uGMRT. The Figure is adapted from \cite{2024MNRAS.529.3372E}.}
    \label{fig:pssph}
\end{figure}

\cite{2023MNRAS.525.3439E} removed the foregrounds from the measured $C_\ell(\Delta\nu)$ to access the largest scales. The key idea here is that the 21-cm signal is localized in $C_\ell(\Delta\nu)$ due to its faster frequency dependence, and therefore it is ideal to remove the foregrounds from $C_\ell(\Delta\nu)$, rather than from visibilities. Later on, \cite{2024MNRAS.529.3372E} applied the method to a large frequency bandwidth $394-494 \, {\rm MHz}$  $(z = 1.9 - 2.6)$, producing the $2\sigma$ upper limit $\Delta^2(k) \leq (4.68)^2 \, \rm{mK^2}$ at $k=0.219\,\rm{Mpc^{-1}}$ which is nearly $15$ times tighter than earlier limits obtained from a smaller bandwidth ($24.4 \, {\rm MHz}$)  of the same data. The upper limit on $[\Omega_{\hi{}} b_{\hi{}}] \leq 1.01 \times 10^{-2}$ is found to be within an order of magnitude of the value expected from the 21-cm signal at these redshifts. \cref{fig:pssph} shows the mean-squared brightness temperature fluctuations $\Delta^2(k)$ obtained from the uGMRT Band 3 data, demonstrating the systematic progress towards a 21-cm IM at $z \sim 2.2$ with uGMRT. An additional 50 hours of observations of the ELAIS-N1 field have since been made and are currently being analyzed to improve the upper limits.

\subsection{Preliminary SKA-Mid Simulation}
\label{sec:prelim_ska_mid}

The techniques developed for the uGMRT can be extended for SKA-Mid. To demonstrate this, we present a preliminary simulation using the SKA-Mid AA* configuration. We consider only the subset of antennas from SKA-Mid AA* that provide coverage around 450~MHz ($z\sim2.2$). This results in 79 numbers of 15-m antennas from the full array. We consider $932$ frequency channels, with a channel width of $\Delta\nu = 53.76\,{\rm kHz}$, resulting in $\approx50\,{\rm MHz}$ bandwidth. We set 6 hours of  observation tracking a target field that transits the SKA-Mid site on 16th June 2025. For comparison, we also simulate 50 hours of observation of the ELAIS-N1 field with uGMRT Band 3 data, which is the data volume currently being analyzed by the uGMRT team.  

The simulated visibility data is modeled as Gaussian random variables with zero mean and a standard deviation anticipated for SKA-Mid \footnote{\url{https://www.skao.int/sites/default/files/documents/SKAO-TEL-0000818-V2_SKA1_Science_Performance.pdf} (Table~6)}. Specifically, we adopt $A_{\rm eff}/T_{\rm sys} = 3.45\,{\rm m^2\,K^{-1}}$ for SKA-Mid at 450~MHz, whereas for uGMRT, we use the standard deviation predicted from existing observations \citep{2024MNRAS.529.3372E}.
For both SKA-Mid and uGMRT configurations, 30 realizations of noise-only simulations are used to produce the results. All the simulated datasets are analysed using the Wideband-TGE formalism as presented in \cite{2024MNRAS.529.3372E}. The power spectrum analysis is restricted to the baselines shorter than 1~k$\lambda$. 

\begin{figure}
    \centering
    \includegraphics[width=0.95\linewidth]{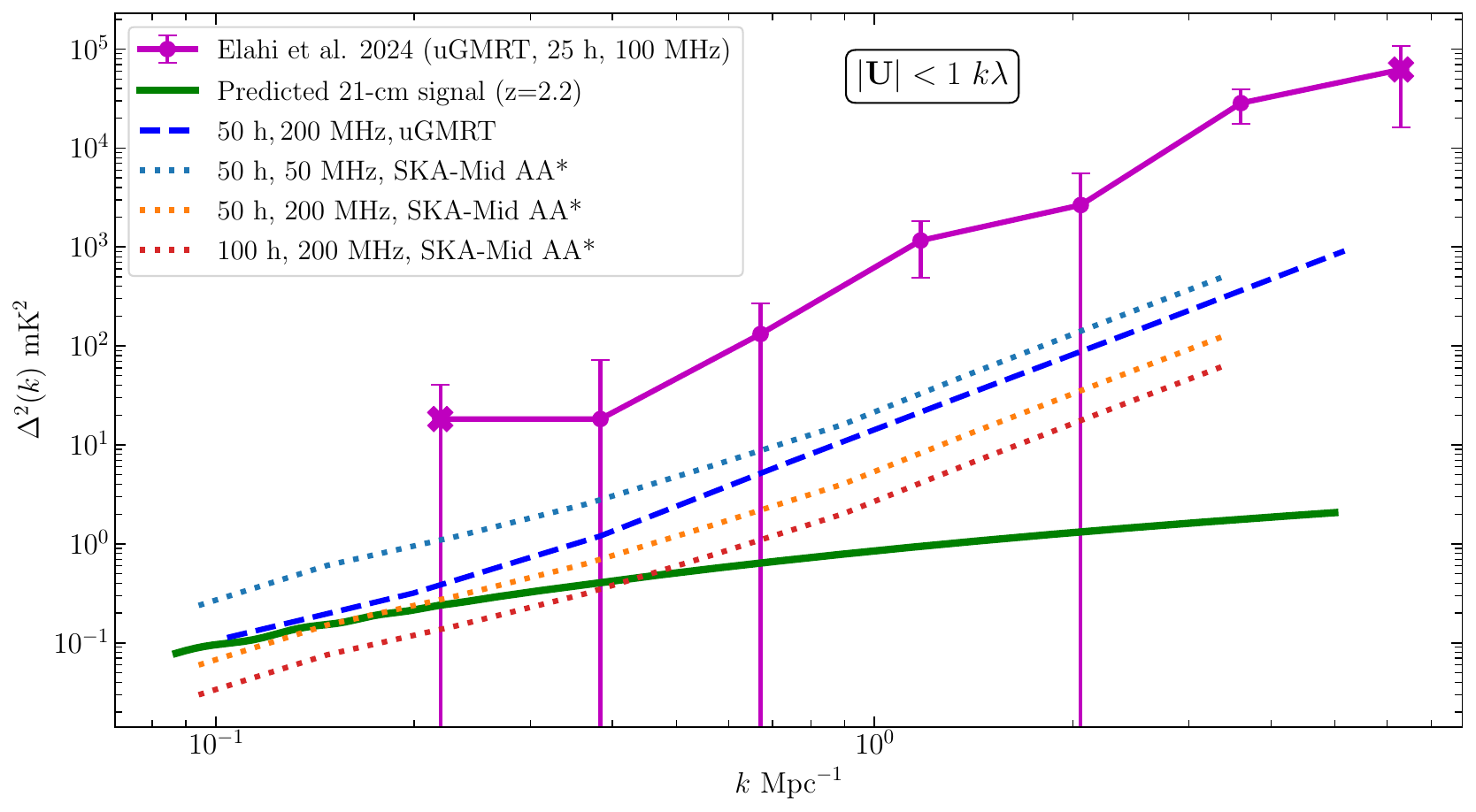}
    \caption{Predictions with SKA-Mid AA* and uGMRT using the noise-only simulations described in Section~\ref{sec:prelim_ska_mid}. Predicted 21-cm signal, and the possibility of its detection with uGMRT and SKA-Mid, using the specific techniques presented in \cite{2024MNRAS.529.3372E}, are shown. The dashed curves show the predicted noise level for different configurations described in the legend. The tightest measurements from uGMRT is shown by the solid magenta points as a reference.}
    \label{fig:pssph_ska}
\end{figure} 

\cref{fig:pssph_ska} shows the predictions with the SKA-Mid along with uGMRT. The upper limits obtained in \cite{2024MNRAS.529.3372E} from 25~h observation over 100~MHz bandwidth are shown as a reference. The ongoing 50~h of on-source uGMRT data promise very close to the detection of the predicted 21-cm signal, although more observations will be required for a successful measurement. Considering SKA-Mid AA*, 50~h of observation will be sufficient to make a detection at length scale $k\sim0.1\,{\rm Mpc}^{-1}$ provided one performs a wideband ($\sim 200$~MHz bandwidth) analysis. Twice the observation time with SKA-Mid AA* holds the possibility for a detection over the length scales of  $0.1<k<0.4\,{\rm Mpc}^{-1}$. We caution that the simulations considered here are noise-only, and they do not include 21-cm signal and foregrounds. Furthermore, the specific simulation and analysis technique are tailored to mimic IM experiments conducted with the uGMRT, and therefore do not necessarily reflect the design specifications or observational strategies of the forthcoming SKA-Mid surveys.

\subsection{Key lessons:}

\begin{itemize}

    \item \textbf{Foreground removal from MAPS:} The IM signal is localized in MAPS $C_\ell(\Delta\nu)$, and therefore it is beneficial to remove the foregrounds from $C_\ell(\Delta\nu)$, instead of visibilities. 
    \item \textbf{Tapering:} Tapering off the sidelobe response can mitigate the oscillatory features in MAPS, thereby it helps in modeling and removing the foregrounds from the MAPS. 
    \item \textbf{$uv$ density:} A dense $uv$ coverage not only boosts sensitivity, but also improves the convolution that is required for a successful tapering of the primary beam. 
    \item \textbf{Dealing with the inevitable missing channels:} The choice of algorithm to handle missing frequency channels is important. Estimating the MAPS $C_\ell(\Delta\nu)$ utilizing the ergodic nature of the 21-cm signal naturally takes care of the flagging \textit{before} doing a Fourier transform.
    \item \textbf{Cross is better:} A cross-correlation between two polarizations mitigates several systematics such as noise-bias, polarized foregrounds, and calibration errors. 
    \item \textbf{Wideband:} A wide bandwidth gives the leverage of improving the SNR once foregrounds are removed from the data. 
\end{itemize}

\section{HIRAX}
\label{sec:hirax}

The Hydrogen Intensity and Real-time Analysis eXperiment \citep[HIRAX;][]{2022JATIS...8a1019C,2016SPIE.9906E..5XN} is a compact radio interferometer array which, at the time of publication, is in the early deployment and pre-commissioning phase. HIRAX will be co-located with SKA-Mid in the South African Radio Astronomy Observatory (SARAO) Karoo Site, in particular on the Swartfontein site $30^\circ 41' 47''$~S, $21^\circ 34' 20''$~E, 12~km from the core of SKA-Mid/MeerKAT. While HIRAX has a range of science goals including detecting and characterising radio transients and \hi{} absorbers, most of its design considerations are tailored towards a high-sensitivity IM survey of the \hi{} line, specifically redshifted into the post-reionization epoch. The telescope will be sensitive to the frequency range 400--800~MHz which corresponds to this line redshifted to $z=0.78-2.55$. The ultimate goal of the project is to comprise an array of 1024 6~m parabolic dishes which are being deployed in stages to the Karoo site. Each dish is re-pointable in elevation, allowing for a large survey of $15\,000$~square degrees of the southern sky.

Fisher forecasts for a telescope with the sensitivity of HIRAX indicate the potential from tight cosmological constraints, including a dark energy equation of state figure of merit of 60 for HIRAX-256 and 360 for HIRAX-1024~\citep{2022JATIS...8a1019C}. Despite the expected statistical power of such a survey, foreground leakage from instrumental systematics remain a major concern for making a high significance detection of the cosmological signal and extracting robust cosmological constraints. The project has focused specifically on mitigating potential systematics as much as possible in the design and commissioning phase of the array. The feeds are placed deep within the dish, with corresponding focal ratio of 0.21, trading sensitivity for reduced cross-coupling effects. The project has also developed tight, simulation informed, requirements~\citep{2020SPIE11445E..5OS} on the manufacturing tolerances of the dish-feed system, with a particular focus on redundancy and therefore repeatability of the manufacturing process\footnote{The HIRAX telescope mechanical assembly requirements document is publicly available at \href{https://hirax.ukzn.ac.za/wp-content/uploads/2021/12/HIRAX\_REQ001\_002\_V1\_Baselined-signed.pdf}{HIRAX\_REQ001\_002\_V1\_Baselined-signed.pdf}}. Using methods evolved from those described in~\citet{2022SPIE12182E..0FI}, the realisation of these requirements has been closely monitored by the science and engineering teams throughout the process of developing the dish manufacturing procedures as well as during the deployment of individual array elements. Simulations of the primary beam and the degree to which it will be required to be modelled have also been studied~\citep{HIRAXSampathApJ}, with further work ongoing to characterize the effect of expected design imperfections. The HIRAX signal chain has been extensively prototyped in the lab~\citep{2020SPIE11445E..2ZK,2022SPIE12182E..25K} and at the HIRAX prototype locations at the Hartebeesthoek Radio Astronomy Observatory (HartRAO) and the SARAO Klerefontein Support Base. These measurements have resulted in significant refinements in the design of the HIRAX feed and signal chain.

\subsection{Current Status}
As of the fourth quarter of 2025, HIRAX is currently in the process of deploying dishes to the Swartfontein site with the local manufacturing facility ramping up production to the expected production cadence of 3 dishes per week at the current facility. The currently planned site development, including foundations for 128 dishes, has been completed and dishes that are currently being delivered to site will be instrumented and commissioned in early 2026. The on-site computing facility including the F-Engine, correlator and science data processor will additionally be installed at the site on a similar timescale.

\section{Murriyang (Parkes)}
\label{sec:Murriyang}

\citet{2009MNRAS.394L...6P} reported the first detection of cosmic structure using 21-cm maps from the cross-correlation of HIPASS \citep{2001MNRAS.322..486B} with the 6dF Galaxy Survey \citep[6dFGS;][]{2009MNRAS.399..683J}, although the correlation function was only able to be measured at low redshifts ($z<0.04$) and over a small range of separations ($<3$~Mpc). Deeper 21-cm intensity maps were made by \citet{2013MNRAS.433.1398D} in the SGP region of the 2dF galaxy redshift survey \citep[2dFGRS;][]{2001MNRAS.328.1039C} for the purposes of examining the \hi{} evolution of galaxies over the redshift range $0.04<z<0.13$) using the method of stacking. More extended SGP and NGP intensity maps were analysed by \citet{2018MNRAS.476.3382A} who reported  a $5.7\sigma$ detection of the cross-power spectrum at $z \sim 0.08$. Further analysis of these maps was used for filament and halo stacking \citep{2019MNRAS.489..385T, 2020MNRAS.498.5916T}. Brightness temperature limits $T_B<10\mu$K were reported.

The above results were all based on data taken with the 13-beam multibeam receiver \citep{1996PASA...13..243S}, which was the precursor of similar systems built by CSIRO for the Arecibo and FAST telescopes. However, \citet{Li_2021} used an uncooled ASKAP phased array feed at lower frequencies to conduct a pilot IM survey. This showed there were sufficient RFI-free gaps at higher redshifts and, despite only having 17 beams and a poor receiver temperature, the survey reported a low-significance detection in four WiggleZ fields at $z\approx 0.76$.

Future IM goals with Murriyang are to conduct pilot surveys with the new cryogenic phased array feed (cryoPAF) at redshifts $0 < z <1$ to assess the likelihood of being able to conduct cross- and auto-correlation surveys, including measurements of BAOs and power spectrum turnover. The Murriyang cryoPAF is potentially a powerful IM instrument with 72 independent beams, excellent system temperature (<20 K), 600 MHz of instantaneous bandwidth, low chromaticity, and an angular resolution which is well-matched to cosmological studies. Commissioning results are reported by \citet{2026PASA...43...57S}.

\section{Tianlai}
\label{sec:tianlai}

The Tianlai experiment is a Chinese pathfinder project specifically designed to develop and test key technologies for 21-cm IM in the post-reionization era \citep{2012IJMPS..12..256C,2015ApJ...798...40X,2020SCPMA..6329862L,2021MNRAS.506.3455W}. Located at the Hongliuxia radio-quiet site in Xinjiang (latitude: $44.15^\circ$N, longitude: $91.80^\circ$E), it features two complementary radio interferometer arrays: a cylinder pathfinder array and a dish pathfinder array \citep{2020SCPMA..6329862L,2021MNRAS.506.3455W}. This dual-array approach allows for comparative testing of different methodologies for large-scale structure surveys using \hi{}.

\subsection{Cylinder Pathfinder Array}
The Tianlai cylinder array consists of three fixed parabolic cylinder reflectors, each measuring 15~m (E-W) $\times$ 40~m (N-S) and aligned north-south \citep{2020SCPMA..6329862L}. This configuration provides a wide field of view (FoV) in the north-south direction ($\gtrsim \pm 60^\circ$ from zenith), while the reflectors focus signals in the east-west direction. The array operates in drift-scan mode, surveying the northern celestial hemisphere as the Earth rotates. The current system operates in the 685--810~MHz frequency band ($z \sim 0.775$--$1.029$).

A distinctive feature is the deployment of 96 dual-linear polarization feeds distributed unevenly across the three cylinders (31, 32, and 33 feeds respectively) with slightly different spacings (41.33~cm, 40.00~cm, and 38.75~cm). This design effectively suppresses grating lobes by breaking the degeneracy in signal arrival times from different directions \citep{2016MNRAS.461.1950Z,2016RAA....16..158Z}. Each polarization channel is designated by its cylinder (A, B, C), feed number, and polarization (X for N-S, Y for E-W), forming a total of 192 signal outputs.

The initial digital correlator provided visibilities with $\sim$1-second integration, suitable for sky map reconstruction and power spectrum estimation for IM purposes \citep{2021A&C....3400439Z}. More recently, the array has been equipped with a digital beamforming backend \citep{2024RAA....24h5010Y}, enabling high-time-resolution beamforming and opening new capabilities for blind searches of fast radio bursts (FRBs) concurrently with cosmological observations.

\subsection{Dish Pathfinder Array}
The Tianlai dish pathfinder array comprises 16 on-axis parabolic dishes, each 6~m in diameter, arranged in a near-hexagonal configuration \citep{2021MNRAS.506.3455W,2016MNRAS.461.1950Z,2022MNRAS.517.4637P}. The dishes are mounted on alt-azimuth pedestals and observe in the same frequency band (685--810~MHz) as the cylinder array. With an average system equivalent flux density (SEFD) of $\sim$14.15~kJy and a field of view of $\sim19.6\,\deg^2$, this array provides complementary interferometric measurements.

Similar to the cylinder array, the dish array originally operated with a correlator producing $\sim$1-second integrated visibilities for IM. The recent addition of a high-speed digital backend now enables millisecond-timescale observations, making it suitable for transient searches like FRBs \citep{2022RAA....22l5007Y}.

\subsection{Science Goals and Status}
The primary scientific goal of the Tianlai project is to measure the large-scale structure of the Universe through 21~cm IM and constrain cosmological parameters, particularly the dark energy equation of state. The experiment faces the key challenge of separating the weak cosmological signal from foregrounds that are 4--5 orders of magnitude brighter, requiring sophisticated analysis techniques \citep{2020PASP..132f2001L}.

Both arrays have been operational since 2016. Their performance characteristics are detailed in \citet{2020SCPMA..6329862L} for the cylinder array and \citet{2021MNRAS.506.3455W} for the dish array, and have been studied using a unmanned aerial vehicle (UAV) \citep{2021IAPM...63f..98Z,2026RAA....26b5001L}. The performance of the arrays has been studied with simulations \citep{2023RAA....23j5008Y}.
Complicated effects arising from reflections and cross-couplings have been observed and studied \citep{2021RAA....21...59L,2022RAA....22f5020S,2024JAI....1350002K}.
The Tianlai project demonstrates critical technological developments for future large-scale IM surveys, including array design, real-time processing systems, and advanced data analysis pipelines for both cosmological mapping and transient detection.

\section{Conclusions}
\label{sec:Conclusions}

The 21-cm intensity mapping (IM) technique is a highly promising new probe for large-scale structure cosmology. It fills the important role of providing a quasi-independent cross-check on galaxy surveys, which is becoming increasingly important as new tensions in the inferred values of cosmological parameters emerge, including potentially paradigm-breaking observations of time-evolving dark energy by DESI \citep{2024JCAP...10..048C, 2025PhRvD.111b3532L, 2025PhRvD.112h3515A}. SKA-Mid will have the raw sensitivity to perform wide and deep surveys over exceptionally large cosmological volumes, extending as far as $z = 3$.

The pathfinder observations reviewed in this chapter have paved the way for SKA-Mid IM surveys, which will potentially be SKAO's major contribution to large-scale structure cosmology in the foreseeable future. 
There are several detections of the cosmological signal in the cross-spectrum (or via the stacking technique) with optical galaxy surveys, which are particularly useful for suppressing residual systematics in the 21-cm data. These have already shown that valuable cosmological and astrophysical measurements can be made with an SKA-Mid 21-cm IM surveys, as its greatly increased volume and redshift range would permit more and larger cross-correlation analyses using existing and near-future galaxy surveys from DESI, 4MOST, \textit{Euclid}, and others. In this sense, the pathfinder instruments have already played their part in establishing the scientific viability of this observational probe. The next step for cross-correlation detections is to observe the baryon acoustic oscillation (BAO) feature---a roughly 5\% effect on the power spectrum that is a key distance indicator in cosmology.

What remains to be demonstrated is the ability to detect the 21-cm signal in autocorrelation however (i.e.\ without relying on external survey data from galaxy surveys or otherwise). This is an important capability, as it allows a truly independent cross-check of other large-scale structure surveys, and allows the 21-cm observations to be extended to redshifts and distance scales that are not yet possible to probe with large galaxy surveys. This will require further advances in data analysis methods, but as the various teams' knowledge of their instruments improves and more data are collected, it seems only a matter of time before a good-quality auto-correlation detection will be confirmed.

As a final point, we acknowledge the expected timeline of SKA-Mid 21-cm IM observations. With the wide-area mapping mode not due for commissioning until 2034, and a survey of sufficiently large area (i.e.\ at least $5\,000\,\deg^2$) likely to need at least $5\,000$ hours of observing time---a large undertaking for what is likely to be a heavily over-subscribed key science programme---we anticipate that a cosmology-grade survey data will not be available until 2040 or so. An additional generation of ``Stage V'' galaxy redshift surveys may have entered operation by then, with concepts such as MegaMapper able to extend observations into the range $2 \lesssim z \lesssim 5$ over around $15\,000\,\deg^2$. The LSST 10-year photometric survey would have long since completed, as would the full \textit{Euclid} galaxy redshift survey. This leaves a period of around 15 years for the pathfinder telescopes to continue their observations and develop the 21-cm IM technique---to make the groundbreaking BAO and autocorrelation detections that will propel 21-cm IM into a top-level cosmological observable.

\section*{Author List Ordering}
The lead author coordinated the preparation of the chapter and contributed to the writing of one of its sections. All other authors contributed to writing different sections, and they are listed in alphabetical order by their last names.

\bibliographystyle{abbrvnat-maxbibnames4}
\bibliography{chapter} 

\end{document}